# Machine Learning and Artificial Intelligence-Driven Multi-Scale Modeling for High Burnup Accident-Tolerant Fuels for Light Water-Based SMR Applications

Md. Shamim Hassan[1], Abid Hossain Khan[1], Richa Verma[2], Dinesh Kumar[3], Kazuma Kobayashi[4], Shoaib Usman[4], Syed Alam[4*]

[1]Institute of Nuclear Power Engineering, Bangladesh University of Engineering and Technology, Dhaka 1000, Bangladesh

[2]Department of Electrical Engineering, Indian Institute of Technology Delhi, Delhi 110016, India

[3]Department of Mechanical Engineering, University of Bristol, Bristol BS8 1TR, UK

[4]Department of Nuclear Engineering and Radiation Science, Missouri University of Science and Technology, Rolla, MO 65409, USA

*Corresponding author: Syed Alam (alams@mst.edu)

**Abstract**

The concept of small modular reactor has changed the outlook for tackling future energy crises. This new reactor technology is very promising considering its lower investment requirements, modularity, design simplicity, and enhanced safety features. The application of artificial intelligence-driven multi-scale modeling (neutronics, thermal hydraulics, fuel performance, etc.) incorporating Digital Twin and associated uncertainties in the research of small modular reactors is a recent concept. In this work, a comprehensive study is conducted on the multiscale modeling of accident-tolerant fuels. The application of these fuels in the light water-based small modular reactors is explored. This chapter also focuses on the application of machine learning and artificial intelligence in the design optimization, control, and monitoring of small modular reactors. Finally, a brief assessment of the research gap on the application of artificial intelligence to the development of high burnup composite accident-tolerant fuels is provided. Necessary actions to fulfill these gaps are also discussed.

## 1. Introduction

Being one of the lowest carbon-emitting sources of power (Wang *et al.*, 2011), the nuclear power industry has become one of the leading global energy sources around the world. From 1950 to 2022, this industry extended its markets in more than thirty countries, accounting for around 10% of the world's power from 440 nuclear reactors (Hore-Lacy, 2010). However, the industry is experiencing slow growth after the events like Chernobyl disaster and the Fukushima disaster.

Although the enhanced safety features of new Gen III+ designs have mitigated the phobia of a severe accident in a nuclear power plant to some extent (Taylor, Dessai and de Bruin, 2014; Sun, Zhu and Meng, 2016), the nuclear era is far from being reincarnated. This is mainly because of the increased investment costs associated with these extended safety systems; very few investors are willing to invest such a huge amount of money on a project that can shut down prematurely if something goes wrong. Therefore, the nuclear community is constantly in search of a safer yet cost-effective substitute of the large-scale Gen III+ nuclear power plant (NPP) designs.

Many leading nations in the nuclear power industry have appeared intrigued by the concept of small modular reactors (SMRs) lately. As a result, new and innovative research projects have been initiated with the collaboration of scientists from top nuclear research laboratories across the globe. Some countries have already planned to install them (Ricotti and Fomin, 2020). On the other side of the table, some countries don't see the commercial viability of SMRs because of its lower efficiency and electrical power output compared to the ordinary light water reactors (LWRs) (Ingersoll, 2010). Bingham and Mancini (2014) mentioned in his research paper that the installation of SMRs is suitable for an installed capacity of 1 to 3 $GW_e$. They also opined that SMRs have the potential of creating new employment opportunities in the nuclear industry. He compared the advantage of the small modular reactor with base-load technology after assessing their life cycles. However, they did acknowledge that the economic and social boundaries would be a major concern for the deployment of SMRs (Bingham and Mancini, 2014).

As an aftermath of the Fukushima disaster, serious questions were raised regarding the fuel performance of light water reactors (LWRs). In the event of a serious nuclear accident, a significant amount of radioactivity could leak into the environment. It became clear during the Fukushima disaster that hydrogen explosions and the release of radionuclides could result in significant health issues. (Koo *et al.*, 2014). Due to the damage of the nuclear fuel rods, which contain fuel pellets and fuel cladding, these two problems i.e., hydrogen explosions and radionuclide release may occur. When exposed to a high-temperature steam environment, Zr-based alloys rapidly oxidize in accident conditions. Hydrogen is generated, resulting in explosions (Nikulina, 2004). The cladding of Zr-based alloys also balloons and opens in an accident, releasing radionuclides into the environment (Kim *et al.*, 2016). To enhance nuclear power plant reliability and safety, fuel claddings are required to maintain their inherent mechanical properties under normal and accident conditions. In addition, it is expected that the fuel pellet itself will prevent release of radioactive fission products during a serious accident. Therefore, recent LWR research is focused primarily on the development of accident-tolerant fuels (ATFs) (Carmack *et al.*, 2013; Kim *et al.*, 2016).

The ATFs are characterized by their innovative design and material that prevent or delay radionuclides from being released during an accident (Montgomery *et al.*, 2013; Kim *et al.*, 2016). While ATFs are expected to tolerate a considerable period of coolant loss compared to conventional $UO_2$-Zr alloy fuels, they should also provide satisfactory performance during normal operations and transient conditions. In addition, they should improve fuel safety for events that are beyond the design basis (Koo *et al.*, 2014). In other words, they must have more oxidation resistance capability with strong mechanical strength under accident conditions (Carmack *et al.*, 2013). To diminish the outcome of the accident, various ATF concepts are being explored (Koo *et al.*, 2014).

As mentioned earlier, the primary objective of ATF is to increase the thermal conductivity and contain the fission products inside the fuel pellets. Moreover, according to the ATF concept, it must be ensured that the density of the uranium is high enough to compensate the burnup or cycle length. (Ray, Johnson and Lahoda, 2013). Not only fuel pellets but also fuel cladding is the main components of ATF. Ceramic composite claddings are being studied for use in ATFs (Kim, Kim and Park, 2013). No-Zr-alloy cladding are also being considered (Terrani, Zinkle and Snead, 2014). Researchers are working on the development of Mo–Zr claddings and iron-based alloy for the midterm application in the reactor (Cheng *et al.*, 2013; Terrani, Zinkle and Snead, 2014). Different types of cladding coating systems ease the production of ATFs (Idarraga-Trujillo *et al.*, 2013). Finally, silicon carbide (SiC) is appraised for long-term application in the reactor (Stempien *et al.*, 2013). During the fabrication of ATF, several factors must be kept in mind such as technical challenges, economics, development schedule, and nonetheless the safety (Carmack *et al.*, 2013).

After installation and commissioning of any power plant, its proper operation and control are the most crucial points to think about. There are numerous events where the accident in a power plant is initiated due to human error rather than malfunction of the plant equipment. Therefore, the researchers are suggesting autonomous control systems for operating the nuclear power plants. Since these control systems must work satisfactorily under transient conditions, machine learning (ML) has recently become the focus of research in this field. Richard T. Wood (2017) believed that artificial intelligence (AI) can be used for controlling power plants to maintain safe and stable operation (Wood, Upadhyaya and Floyd, 2017). Artificial neural network (ANN) is the skeleton of computational intelligence techniques. Because of its diversity, ANN can be used to solve different nuclear engineering problems (Manic and Sabharwall, 2011). Computational intelligence (CI) has been applied in the nuclear industry for many years, although the lack of real-life data of plant transient and the Blackbox nature of AI has been a matter of concern in many situations (Suman, 2021). In his work, Bernard (1989) explored the applicability of ANN for nuclear system. Uhrig (1991) wrote an encapsulated summary of neural network applications in nuclear reactor monitoring. Using AI, Reifman (1997) investigated mechanisms of detecting and locating faults in nuclear equipment. Uhrig and Tsoukalas (1999) reviewed how to use fuzzy logic, neural networks, and genetic algorithms to operate, monitor, and diagnose nuclear reactors. In a latter review articles by Uhrig and Wines (2005), it was pointed out that there was a gradual shift in the focus towards implementation of CI in nuclear power plant design and optimization. Recently, research is being carried out on designing methods for implementing AI techniques for next-generation reactors and space reactors, which are under development (Uhrig and Hines, 2005). advanced data-driven AI approaches to transient detection in reactors is also becoming very common now-a-days (Moshkbar-Bakhshayesh and Ghofrani, 2013). Furthermore, digital twin framework using constructive AI/ML surrogate model for ATF based has been proposed by General Atomics Electromagnetic Systems (Jacobsen, 2022) in collaboration with Idaho National Lab (INL) and Los Alamos National Lab (LANL).

From the above literature survey, it is evident that SMRs, ML and ATFs are three major topics of research in recent times, but they are yet to cross paths. This chapter focused on the latest progress in the research on machine learning and artificial intelligence for designing ATFs for the LWR-based SMRs. After the informative introductory part, brief descriptions of ATF, ML and AI are given. Also, the application of ATFs in light water-cooled SMRs is summarized.

Then, the current trend in the use of artificial intelligence in the field of nuclear engineering is presented. Finally, the prospect of using Artificial Neural Network (ANN) in developing high burnup composite ATFs for light-water cooled SMRs is explored. Finally, research gap and scope for future studies are also discussed.

## 2. Accident tolerant fuels for light water reactors

Accident-tolerant fuel (ATF) is a challenging area of research in the field of nuclear engineering. Normally, the core damage frequency of any reactor is once in a million years (Purba *et al.*, 2020). But when it happens, the consequences are catastrophic. Thus, precautions are required to prevent core damage in case of any form of accident, design basis or beyond design basis. Currently, $UO_2$-Zircaloy based fuel is the most familiar fuel in the nuclear industry. However, there properties make them vulnerable to severe accidents (Alam, Goodwin and Parks, 2019a; Almutairi *et al.*, 2022).

The temperatures of the fuel and the moderator increase if a leak occurs in the coolant pipe or if the pump stops working due to an accident in a conventional light water reactor (LWR). The physical components, as well as chemical debasement, occur when the fuel temperature exceeds 800°C. For temperatures between 700°C to 1200°C, there is a probability of damage of the fuel rod such as release of radioactive gases, oxidation of the material, and dangerous reactions initiated between the Zr cladding and fuel (Purba *et al.*, 2020). Without an emergency core cooling system, there is a chance of the formation of hydrogen gas. In the presence of oxygen, hydrogen gas is likely to explode. When temperatures rise above 1500°C, there is a greater chance of a core meltdown (Purba *et al.*, 2020). Fig.1 presents the possible damaging events of a typical LWR fuel rod after an accident. Since the next generation reactors are expected to be operational at higher power densities, there limitations must be overcome. Therefore, $UO_2$-Zircaloy based fuel is expected to be replaced by ATFs in future as they can withstand loss of active cooling for a longer period without any damage (Zinkle *et al.*, 2014). The major advantages of ATFs over $UO_2$-Zircaloy based fuel are (Ott, Robb and Wang, 2014; Zinkle *et al.*, 2014):

- Hydrogen generation at a much slower rate
- Improved reaction dynamics with steam
- Superior thermal stability and fission product retention of cladding
- Enhanced mechanical strength and thermal conductivity of fuel

"The Consolidated Appropriations Act, 2012", approved by the Department of Energy, Office of Nuclear Energy, USA (Goldner, 2012), has given notable practical and scientific emphasis on the development of ATFs. Multiple projects were sponsored to increase the safety and security system of the current generation's reactors. These projects aim at evaluation of different ATFs and their potential use in LWRs. The projects also focus on successful implementation of the design upgrades in the commercial nuclear power plants before 2021 (Goldner, 2012).

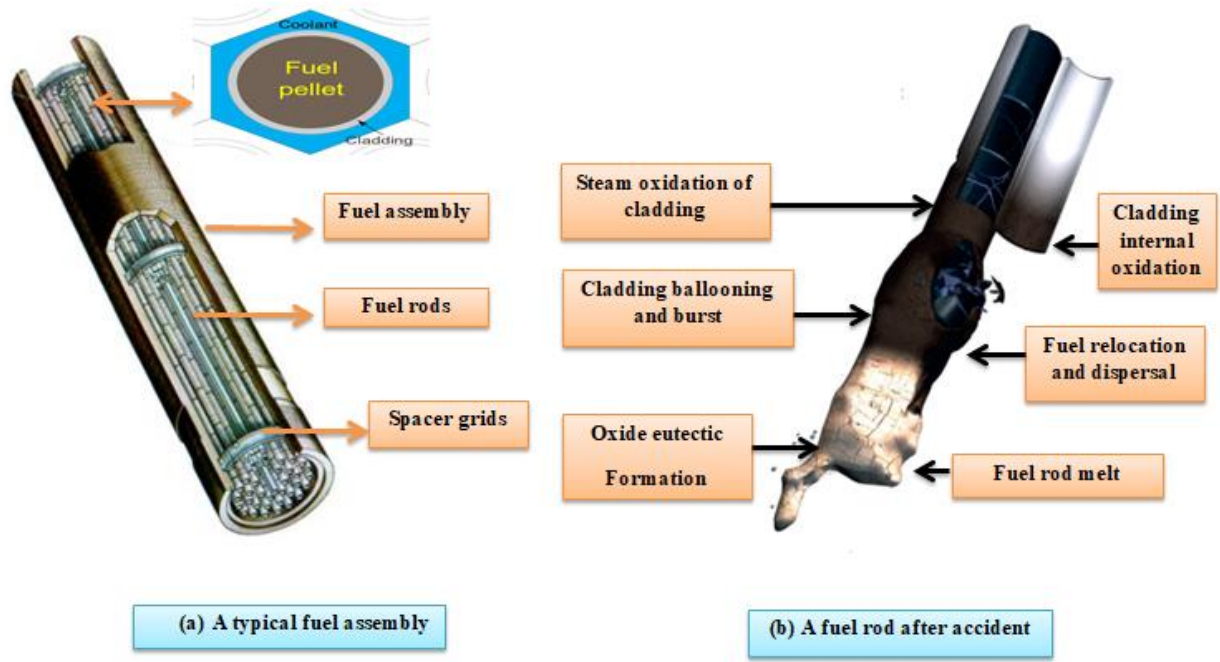

**Fig.1.** (a) A typical fuel assembly, (b) A fuel rod after accident (Zinkle *et al.*, 2014)

Researchers are working on different ATF materials for prospective use in light water reactors (LWRs) (Bragg-Sitton *et al.*, 2016; Spencer *et al.*, 2016; Terrani, 2018; Ebrahimgol, Aghaie and Zolfaghari, 2021). Table 1 summarizes the recent research focused on the properties of ATF materials and their responses to normal and transient conditions. From Table 1, it may be observed that the researchers are considering multiple options; accident tolerant cladding materials with conventional $UO_2$ fuels, improved fuel material with conventional Zr-based alloy or a completely new fuel-cladding system (Chen, He and Yuan, 2020). Numerous studies have observed promising properties in Fully Ceramic Microencapsulated (FCM) (Powers *et al.*, 2013; He *et al.*, 2019) and $U_3Si_2$ (Nelson *et al.*, 2014) fuels. However, a few technical challenges such as higher volumetric swelling, lower melting point, inferior neutronic behavior, etc. compared to $UO_2$ are yet to be resolved (Spencer *et al.*, 2016; Chen, He and Yuan, 2020). Some researchers have also suggested addition of BeO or Mo in $UO_2$ fuels for improved thermal conductivity, although the inclusions may cause changes in the neutronic behavior of the fuel (Chandramouli and Revankar, 2014). On the other hand, high Chromium SS and SiC-based cladding materials are slowly gaining attention of the nuclear community (Brown *et al.*, 2017).

**Table 1.** Summary of recent research works on ATFs for LWRs

| Author(s) | Research Outline | Key Findings |
|---|---|---|
| (Powers *et al.*, 2013) | FCM as alternate fuel in LWRs | 1. FCM fuel can overcome the problems associated UN fuel in LWRs. |
| (Chandramouli and Revankar, 2014) | Thermal modelling and analysis of UO2-BeO fuels | 1. Thermal conductivity is significantly increased due to addition of BeO in $UO_2$ fuel. |
| (Nelson *et al.*, 2014) | U-Si binary as ATF | 1. $U_3Si_2$ has high thermodynamic stability under reactor transient conditions. |
| (Ott, Robb and Wang, 2014) | Analysis of the performance of different ATFs over the full spectrum of accident and beyond design accident conditions | 1. Peak cladding temperature may be reduced up to 75ºC if Zr is replaced by FeCrAl cladding<br>2. Lower peak cladding temperature may be achieved for FCM fuel. |
| (Wu *et al.*, 2015) | Investigation of performance of ATF during accident events | 1. FCM-SiC fuel-cladding system is found to be more accident tolerant than other ATFs for LBLOCA with extended SBO accident. |
| (Spencer *et al.*, 2016) | Properties of different ATFs | 1. Irradiation-induced swelling of UN fuel is very high compared to $UO_2$ and $U_3Si_2$ |
| (Kim *et al.*, 2016) | Progress of ATF in Korean LWRs | 1. Zr-based alloy (Modified) and SiC exhibit potential ATF cladding materials for enhanced safety. |
| (Brown *et al.*, 2017) | Response of ATF cladding materials on reactivity-initiated accidents (RIA) | 1. Different cladding materials (SiC-SiC, Zircaloy, and FeCrAl) have their unique response and neutronic behavior during a RIA. |
| (Karoutas *et al.*, 2018) | Development of ATFs by General Electric, USA | 1. SiC/SiC composite cladding has better accident tolerant properties compared to Zr-based claddings.<br>2. $U_3Si_2$ fuel pellets have shown promising resistance against failure during a Three Mile Island (TMI)-like accident scenario.<br>3. High Cr SS claddings such as FeCrAl or APMT have oxidation resistance up 1400-1500ºC |

### 3. ATF research for LWR-based SMRs

Small modular reactors (SMRs) are advanced Gen IV reactors with improved safety (Aydogan, 2016). There are many differences between the current light water reactor and small modular reactors (SMRs). The two words, "small" and "modular" indicate the main differences between the current light water reactor and SMRs. The name "small" refers to the low electric power output of these reactors, typically below 300MW$_e$. And the name "modular" indicates that these reactors are constructed off-site. Although gas-cooled and metal-cooled SMRs are also being considered, light water cooled SMRs (LW-SMRs) are the most likely reactor technologies to be available commercially in near future (Aydogan, 2016).

In every typical LW-SMRs, the numbers of fuel assemblies are identical, that is 17×17. The basic main differences between the commercial light water reactors and LW-SMRs are in their sizes, lengths of the fuel cycle and the output power capacity. The reactivity control system of LW-SMRs is identical to LWRs. Control rod assemblies are used for controlling the reactivity of the LW-SMRs. Moreover, for controlling the reactivity, soluble and burnable absorbers are also being considered in most designs (Aydogan, 2016). No revolutionary change in the fuel element of commercial LW-SMRs has been proposed either. Table 2 indicates the proposed fuel components in LW-SMRs.

**Table 2.** the comparison of fuel components in LW-SMRs (Aydogan, 2016)

| SMRs | Fuel | Enrichment (%) | Cladding |
|---|---|---|---|
| NuScale | UO$_2$ | 4.95 | Zr-4 or advanced cladding |
| W-SMR | UO$_2$ | <5 | ZIRLO |
| IRIS | UO$_2$ | <5 | Zr Alloy |
| SMART | UO$_2$ | <5 | Zr-4 |
| mPower | UO$_2$ | <5 | Stainless steel |

From Table 2, it is observed that there is no difference in the fueling material of different types of LW-SMRs; all of them are considering conventional UO$_2$ fuel pins. Their enrichment level is about five percent. However, the cladding materials are distinguishable for different LW-SMRs. Nevertheless, almost all of them are Zirconium-based alloys. Only mPower has considered SS cladding. The reason behind choosing the conventional fuel materials for the LW-SMRs is the safety of these novel reactor designs. Since these reactors have lack of real-life data during operational state, the transient behavior of these reactors can't be predicted with sufficient confidence (Aydogan, 2016). Therefore, the manufacturers are going for the proven technologies of their predecessors i.e., the large-scale LWRs currently operational. However, the use of ATF can significantly improve the safety of LW-SMRs (Awan *et al.*, 2018; Tiang and Xiao, 2021). Therefore, researchers are continuously exploring different fuel options for LW-SMRs. Table 3 summarizes some of the latest research on the application of ATFs in LW-SMRs.

**Table 3.** Summary of recent research works on prospective ATFs for LW-SMRs

| Author(s) | Research Outline | Key Findings |
|---|---|---|
| (Awan *et al.*, 2018) | IPWR core with accident tolerant fuel | 1. Low Enriched Uranium Carbide embedded in SiC matrix with FeCrAl cladding has higher fission product retention and lower $H_2$ production<br>2. AT-FCM loaded IPWR has excellent neutronic parameters |
| (Li *et al.*, 2019) | Application of $U_3Si_2$ – FeCrAl ATF in marine SMR | 1. In a marine SMR with $U_3Si_2$ – FeCrAl ATF |
| (Pino-Medina and François, 2021) | NuScale Power core utilizing ATFs with different coating materials from Framatome | 1. Use of $U_3Si_2$ ATF with different coating materials in Nuscale is safe and feasible. |
| (Tiang and Xiao, 2021) | ATF-loaded marine SMRs to assess the long-term reactivity control | 1. Fuel with FeCrAL cladding and $B_4C$ + $Gd_2O_3$ particle-type burnable poison can prevent prompt criticality without the aid of control rods in Russian KLT-40S SMR. |
| (Pourrostam, Talebi and Safarzadeh, 2021) | Accident tolerant cladding for SMART | 1. Use of both SiC and FeCrAl can decrease average fuel temperature and increases integral control rod worth.<br>2. The AT claddings can increase both the fuel and moderator temperature coefficients. |

From Table 3, it may be observed that ATFs have promising performances in LW-SMRs, although the fuel and moderator temperature coefficients may increase depending on the fuel-moderator combination (Pourrostam, Talebi and Safarzadeh, 2021). Nevertheless, the unavailability of sufficient data makes it difficult to come to a definite conclusion. Further research is required to access the prospective use of ATFs in LW-SMRs.

## 4. Machine learning and nuclear power industry

Scientists have worked hard for years to develop intelligent machines, and the pursuit of perfect machines is seemingly never-ending. These machines are programmable and can help accomplish any target goal. Intelligence is the capacity to perceive knowledge. It helps in making decisions based on reasoning and learning. Many researchers think that machine learning (ML) is a crucial part of artificial intelligence (AI) because they help to parse data. It also grasps this data and applies the learning to make decisions. Reinforcement learning, supervised learning, and unsupervised learning are the three main framework of ML, as shown in Fig.2. While these three have their specific areas of application, supervised learning is one of the most adapted ML frameworks. It is employed for the categorization or regression of data. Neural networks, support vector machines, and naïve Bayes are examples of supervised learning (Suman, 2021). The use of supervised learning is dependent on the availability of training data.

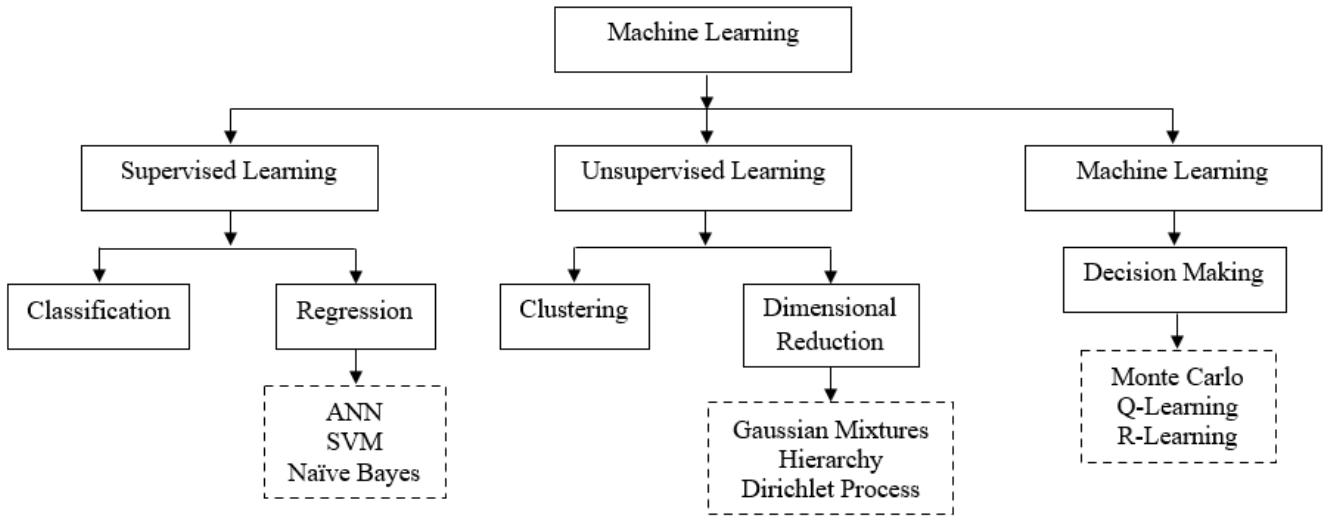

Fig.2: Machine learning classification (Suman, 2021)

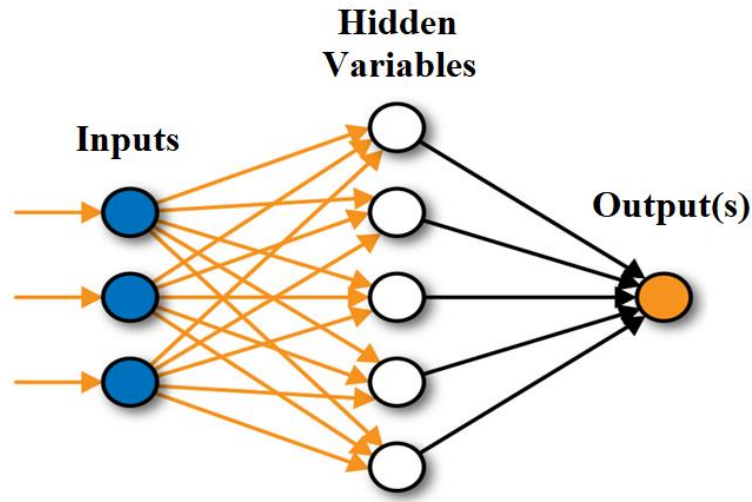

Fig.3: Layers of neural network

Neural network (NN), also known as artificial neural network (ANN), is the most common type of supervised learning. The concept of NN comes from the biological working principle of the animal brain (Wood, Upadhyaya and Floyd, 2017). The structural framework of NN is shown in Fig.3. The input layer is the primary layer of the NN. The hidden layer is the second layer after the input layer. And the output layer is the final stage. A neuron is the key element of the network and connects these layers. The meeting points of the neurons can be referred to as a node. This neural network is trained different learning algorithms. NN can be divided into a few sub-branches. In recent years, researchers are working more on the recurrent neural network, backpropagation neural network, radial basis neural network, and feed-forward neural network because of their major application in the nuclear industry. But a preference is observed for backpropagation neural networks because of their wide application in this sector (Suman, 2021).

The potential use of NN in the field of the nuclear power plant is expanding day by day (Suman, 2021). This has resulted from the nuclear industry's effort to stay on top of technological advances. The research groups of the major manufacturing industries of nuclear reactors are working on developing innovative reactor designs that are intrinsically safe and secured, clean and reputable, inexpensive, workable, and well-grounded. They have changed and improved the potential safety system of the reactor generation wise (Wood, Upadhyaya and Floyd, 2017; Suman, 2021). Likewise, they are putting their effort to enhance the overall capacity and availability factor by reducing the maintenance and operational cost of the current reactors. ANN has emerged as a potential Fastlane to reach these goals (Suman, 2021) [1].

In terms of modeling and controlling the nuclear reactor to forecasting and pattern recognition, the use of the NN is praiseworthy (Wood, Upadhyaya and Floyd, 2017). During any abnormal condition, the operator needs to make a quick decision to handle the situation. NN-based computational system can help the operator tackle this type of situation. It can also reduce the volume of work of the operator as well as assist in problem-solving during any kind of unwanted condition. Further, it can help to build up intelligible resources (Suman, 2021).

To reduce the overall cost, NN may be used to determine the optimum reactor core size, necessary structural change of the core reload design, management of the fuel in the core, and designing of the fuel pellets and lattices, etc. (Uhrig and Hines, 2005; Suman, 2021). NN can enhance the capability of a NPP by diagnosis the fault condition of the plant, monitoring the reactor operational status, detecting the failure of the sensors, actuator, and transducer condition. During any transient condition, NN can help in managing the transient of the power plants by classifying it according to safety criteria depending on its severity (Suman, 2021). In short, NN may be used for optimization, monitoring and control of a NPP with high efficiency.

## 5. Machine learning and AI in nuclear materials research

ML and AI have brought about a revolution in modern research. With this powerful tool, a researcher can make a machine do the tasks on its own that would have needed human intervention otherwise. And the thing is, machines are faster than men. So, artificial intelligence has, to some extent, contributed to the rapid expansion of modern science, engineering and technology. And since nuclear engineering is a complicated and sensitive branch of engineering, no wonder researchers have employed artificial intelligence, especially ANN, in different applications related to this field (Suman, 2021). The material science has gathered a lot of attention in recent times. The future reactor designs ask for advanced materials with better mechanical, chemical and neutronic properties [citation]. Machine learning and artificial intelligence can contribute to the rapid development of this field (Morgan *et al.*, 2022).

AI may be utilized in numerous possible ways in nuclear material research. The most common application is prediction of mechanical properties of the structural materials in a nuclear facility (Sharma, Bhasin and Ghosh, 2010; Sharma *et al.*, 2011; Morgan *et al.*, 2022). The structural components may include Reactor Pressure Vessel, piping systems, containment building, etc. (Morgan *et al.*, 2022). NN is also found to be very efficient in fault detection of spent nuclear fuel (Rossa, Borella and Giani, 2020). Perhaps the most sophisticated application of AI in material science is the modeling of subatomic and interatomic behavior of different materials. The interatomic potential of materials of Gen IV reactors such as molten salts (Lam *et al.*, 2021; Li *et al.*, 2021; Sivaraman *et al.*, 2021), high-grade stainless steels (Min *et al.*, 2021), etc. are being studied with the help of artificial intelligence. AI is also utilized for evaluation and validation of nuclear data (Neudecker *et al.*, 2020).

Nuclear safeguard has become a topic of sufficient debate and discussions within the nuclear community. The international organizations are proactive to prevent proliferation of nuclear materials, especially Plutonium and Uranium-233. AI is being employed at a regular basis in the research related to identification of these Special Nuclear Materials (SNMs). Some of the recent research proposed the use of gamma-spectroscopy (Curtis, 2016; Zhang *et al.*, 2019) for generating data for training the neural network while the others have suggested using a mixed photon-neutron environment (Jinia *et al.*, 2021; Zhang *et al.*, 2021).

## 6. AI-driven research in LWR-based SMRs

The use of AI has already become very common in optimization, monitoring and control of modern Gen III+ reactors (Suman, 2021). The application of NN in improving the design and safety of advanced and Gen IV reactors, including SMRs, is also being explored (Manic and Sabharwall, 2011). In the early part of the century, the use of NN was observed for the thermal-hydraulic design optimization of different components of a NPP (Ridluan, Manic and Tokuhiro, 2009; Manic and Sabharwall, 2011). The recent research works are somewhat more focused on the core optimization, monitoring, transient behavior analysis and control of a nuclear reactor. An example is the research carried out on the SMR test facility of Oregon State University (Gomez Fernandez *et al.*, 2017). It is a multi-disciplinary test facility used to generate data by changing the core configurations. The behavior of the system during loss of coolant scenario is also recorded. Using the generated data, NN was trained so that it can anticipate the best possible way of behaving i.e., initiating the safety sequences with better precision in different scenarios (Gomez Fernandez *et al.*, 2017). The use of NN in designing the core and neutronic parameters of the high conversion small modular reactor (HCSMR) was explored in another study (Janin, 2018). With the trained NN, some key parameters such as the type of fuel, reflector, and the amount of plutonium in the reactor core were optimized using multi-objective optimization techniques.

A nuclear power plant's safety is of prime importance. The safety of a NPP can be compromised due to multiple reasons such as mechanical failure of plant components, fault propagation, power peaking at undesired rate, etc. To ensure plant safety, these events must be predicted and prevented beforehand. AI may be utilized for this purpose. The reactor power transient has a direct relationship to the transient thermo-mechanical load on the structural components, especially Reactor Pressure Vessel (RPV), nozzles and the piping systems. The use of ANN in modeling the behavior of the structural components subjected to transient loading was proposed by Santucho, Chimenti and Duo (2019). A finite element analysis-based thermo-mechanical model was developed by Saeed et al. (2020) that utilized neural networks to diagnose fault in IP-200 SMR. The result of the work suggested that the model is a reasonable strategy for the diagnosis of the fault in NPPs (Saeed *et al.*, 2020). Yao, Wang and Xie (2022) proposed adaptive residual convolutional neural networks (ARCNNs) for SMRs to detect different kinds of faults due to accidents. Chinese lead-based nuclear reactor (CLEAR) was used to generate experimental data for training the neural network (Yao, Wang and Xie, 2022).

In a boron-free SMR, the reactor power level as well as reactivity is controlled only with the help of control rods. Thus, the movement of rod rods has a direct link to the power peaking factor (PPF) of the SMR. Sanchez and dos Santos (2021) demonstrated the use of Support Vector Machine (SVM), a regression model of AI, to anticipate the PPF in a boron-free SMR. In case of a natural circulation cooling-type SMR like NuScale, the main task of the safety systems is to ensure adequate heat transfer from the core. Rahnama and Ansarifar (2021) used an advanced ANN is used to anticipate the neutronic parameters, as well as thermal-hydraulic variables with the $Al_2O_3$-water nanofluid in Nuscale. The change in safety performance of the reactor due to the introduction of the nanofluid was investigated (Rahnama and Ansarifar, 2021).

## 7. Machine Learning and Artificial Intelligence-based Multi-Scale Modeling Incorporating Digital Twin

In order to speed up the deployment of advanced nuclear energy technologies such as advanced ATFs, the Nuclear Energy Advanced Modeling and Simulation (NEAMS) program is a DOE-NE (Office of Nuclear Energy) initiative (NEAMS, 2022a). There are scopes of multi-scale modeling in NEAMS suited for ATF design in terms of neutronics, thermal-hydraulics, fuel performance and structural performance.

### (a) Neutronics Calculation and Optimization:

Generally, NEAM code MPACT can be used for neutronics. MPACT (NEAMS, 2022b) can be used effectively under the MOOSE Framework. DOE Fuel Cycle R&D expresses interest in having a coordinating finding with NEAMS tools and therefore, MPACT code can be utilized. MPACT uses the subgroup method to generate inline cross sections, with depletion performed using the ORIGEN program. In order to obtain the higher burnup, deep learning-based optimization can be performed using lattice geometry optimization (Alam, Goodwin and Parks, 2019b).

***Deep Learning Surrogate Model Optimization:*** The authors are developing a deep learning surrogate model optimization model, similar to the one referred and studied in a 2020 Cambridge PhD thesis (Whyte, 2020; Whyte and Parks, 2021a, 2021b). Because fuel geometry optimization is critical to extending burnup and increasing the power rating, thus enhancing the competitiveness, this optimization receives special attention. Deep Learning Surrogate Model Optimization (SMO) process (Whyte, 2020; Whyte and Parks, 2021b) uses Deep Multi Layer Perceptrons and Convolutional Neural Networks with iterative robust optimization in order to ensure simultaneous improvement of discharge burnup and power density of the reactor (Fig. 4).

Optimization method is as follows: (1) A training set will be generated; (2) Deep learning techniques will be applied to the training set to generate a model; (3) The model will be used to simultaneously optimize power uprating and burnup extension; and (4) The resulting optimized solutions in terms of design spaces will be tested using the original accurate neutronics simulation. Gaussian Process (GP)-based Bayesian optimization can also be implemented for optimization, as shown in Fig.5.

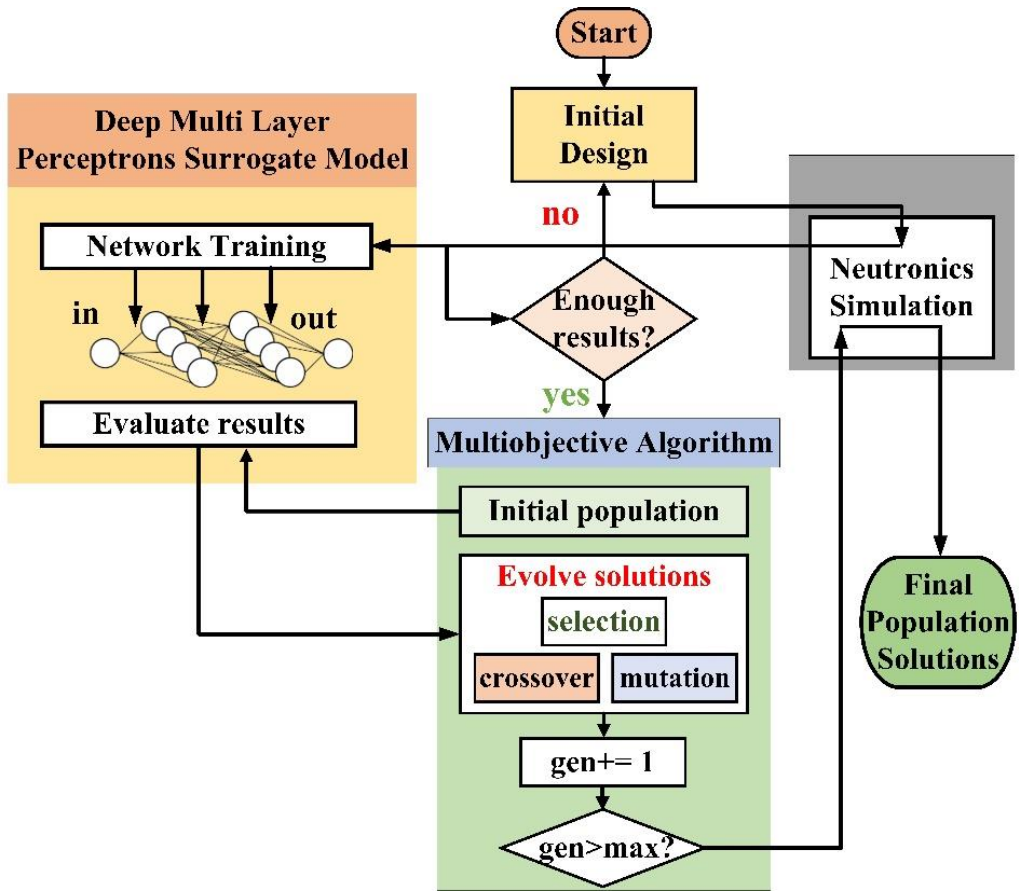

Fig 4. Deep Learning Surrogate Model Optimization Framework (Whyte, 2020; Whyte and Parks, 2021b)

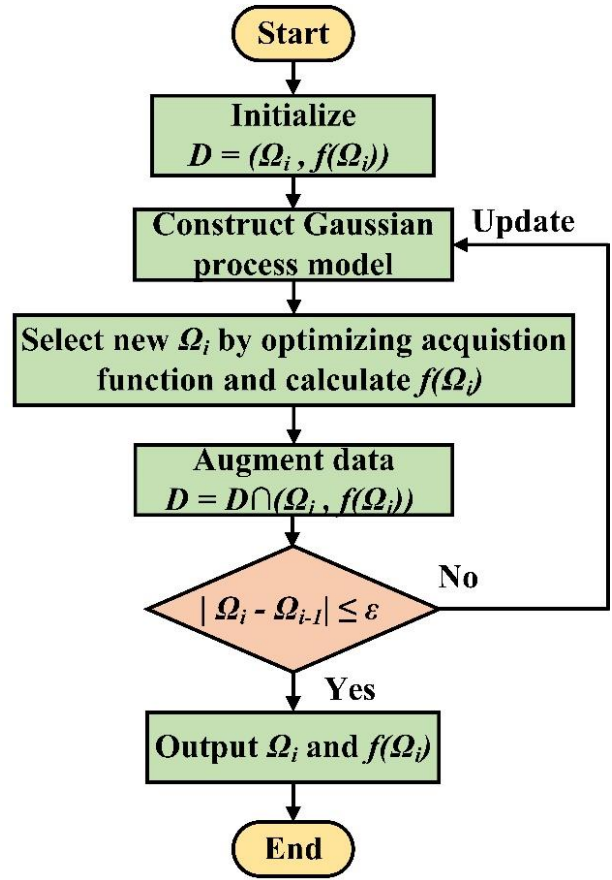

Fig. 5: Bayesian Optimization Process

***Simulated Quantum Annealing (SQA) Optimization:*** Adiabatic quantum computers are used, as is the technique of "quantum annealing," to produce a surrogate model that encodes the heuristics for optimizing burnup and power density. This model can then be used in modern "quantum annealers." Heuristic rules (Galperin, 1995) can be used by a quantum annealer (QA) (Fingerhuth, Babej and Wittek, 2018; Whyte and Parks, 2021a). Quantum adiabatic theory is a method for global optimizing which can be employed in time dependent Hamiltonian. A logical qubit is assigned to each of the assembly's fuel pins for this optimization. Convergence is achieved by converting Ising model connections into actual target QA architectures. The Ising description allows for the connection qubits regardless of configuration (Whyte and Parks, 2021a), which is referred to as "minor embedding". Individual qubit is coupled with two adjacent groups and each group is connected to two adjacent qubits in the box labeled 'Minor graph embedding' in Fig. 6. It is necessary to couple several real qubits together in order to simulate the connectivity of a logical qubit.

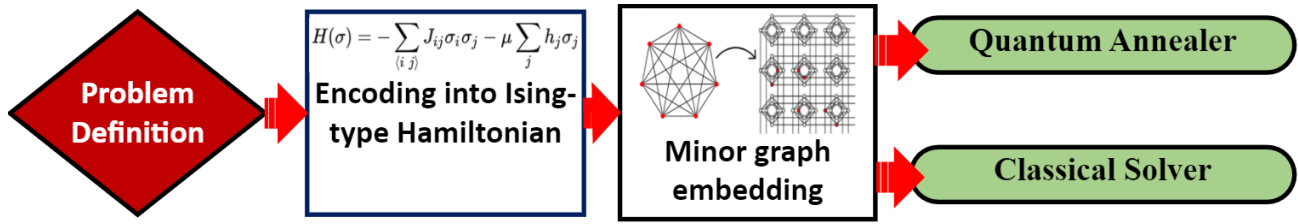

Fig.6: Quantum annealing optimization process (Fingerhuth, Babej and Wittek, 2018; Whyte and Parks, 2021a).

*(b) Thermal-Hydraulics:*

The COBRATF (NEAMS, 2022b) thermal-hydraulics subchannel code has been updated as CTF, and can be used effectively under the MOOSE Framework. It is fully integrated with MPACT neutronics code. As addressed, since DOE Fuel Cycle R&D expresses interest in having a coordinating finding with NEAMS tools and therefore, MPACT/CTF coupled code can be utilized. Solid thermal conduction modeling for tubes, cylinders, and nuclear fuel rods is also available from CTF. Burnup-dependent fuel temperatures are modelled by CTF Fuel, a nuclear fuel rod solver. Also, wide ranges of neutronics and thermal-hydraulics codes can be also coupled for this analysis, as shown in Fig.7 (Alam *et al.*, 2019).

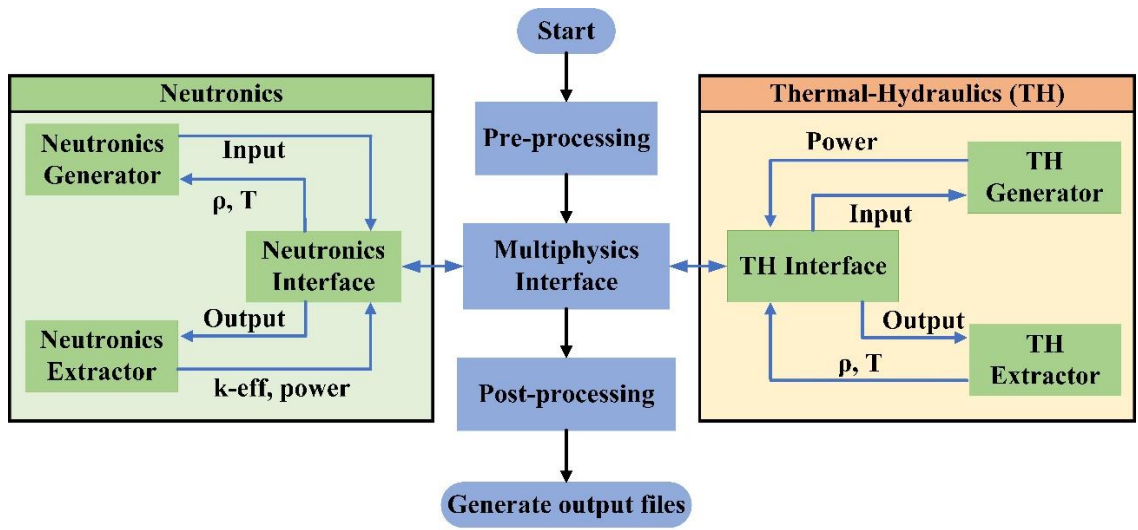

Fig. 7: Coupled Neutronic/Thermal hydraulic Software (Alam *et al.*, 2019)

*(c) Fuel Performance Code:*

BISON (INL, 2022) is a finite element-based nuclear fuel performance code that can be used for ATF with SMR. BISON can perform wide range of fuel performance calculations such as fission gas release, axial fuel swelling, fuel porosity, cladding strain, crack propagation, thermal and irradiation creep, fracture and so on.

BISON is now fully coupled to the mesoscale fuel performance code MARMOT. Because of its foundation in the MOOSE framework, BISON is capable of running on both standard workstations and high-performance computers.

### (d) Digital Twin Surrogate Models:

As referred to our study (Kobayashi *et al.*, 2022), the digital twin framework is proposed to develop using constructive surrogate model. The authors are developing a new surrogate modeling tool for the digital twin system for ATF based on the General Atomics Electromagnetic Systems proposal for the DOE Nuclear Power Program Technologies (Jacobsen, 2022). The method is proposed by George Jacobsen of General Atomics Electromagnetic Systems in collaboration with Idaho National Lab (INL) and Los Alamos National Lab (LANL). This digital twin framework results in various short-term and long-term ATF concepts and their implementation. In the proposed work, a collaboration with the GE team, INL and LANL would be important. This is also important to ensure the developed AI-driven and surrogate assisted digital twin framework is accurate and validating it against experimental data available from literature and industry. "Virtual twin" of ATF system is the goal of this proposed project. According to General Atomics Electromagnetic Systems (Jacobsen, 2022), we are developing digital twin framework following the route (which has also been addressed for hybrid energy system in another study (Khan *et al.*, 2022): (1) The goal is to build a ML/AI-driven surrogate model to understand the response of ATF in LWR-based SMR environment, while using simple equations to keep the model as simple as possible. (2) We can perform surrogate model validation using a more empirical approach by incorporating relevant data and incorporating it into existing ATF models/data. (3) Expand the ATF model to incorporate the ATF surrogate model's behavior. (4) Use data from the system and experiments to figure out and limit model uncertainties for all models. Fig.8. shows the framework (Kochunas and Huan, 2021; Kobayashi *et al.*, 2022) for interactions between digital twin and physical assets.

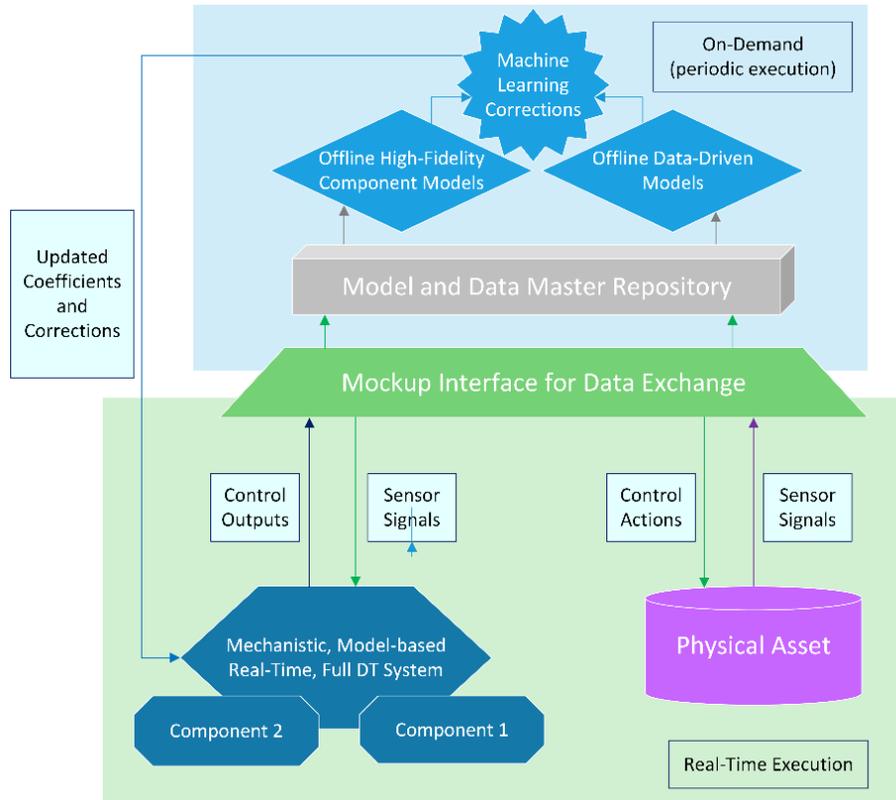

Fig.8. Framework for interactions between digital twin and physical assets (Kochunas and Huan, 2021; Kobayashi *et al.*, 2022)

### *(e) Robustness Analysis Using Machine Learning:*

ATF designs have certain uncertainties associated with its input and model parameters e.g., uncertainties in the thermal-hydraulics parameters, initial condition, boundary condition, etc. Since these uncertainties are propagated to the system response, the influence of these uncertainties must be estimated and described in the final solution of the proposed design to ensure the quality and reliability of the results. To ensure reliable ATF design, it is necessary to provide accurate input information in the models. In order to ensure a robust design, uncertainty quantification (UQ) and sensitivity analysis (SA) (Pepper, Montomoli and Sharma, 2019; Kumar, Alam, Vučinić, *et al.*, 2020; Kumar, Koutsawa, *et al.*, 2020; Kumar *et al.*, 2021, 2022) will be performed to ensure that all the design parameters and their variabilities are within the safety margin. It is also important to consider sensitivity analysis to determine how the proposed design is affected due to the variability in design variables (Kumar *et al.*, 2019; Kumar, Alam, Sjöstrand, *et al.*, 2020).

The authors developed a non-intrusive Polynomial Chaos approach for the uncertainty quantification to understand the variability of a core output parameters with respect to the fluctuations in the core input values for the ATF design for SMR core response, as shown in Fig.9. It would be practical to quantify the 'combined' effect of uncertainties simultaneously to realize the real core condition for the proposed SMR concept. This requires a large number of simulations for UQ and the classical polynomial chaos approach becomes ineffective for this concept. To improve the efficiency of polynomial chaos, a sparse polynomial chaos expansion (SPCE) is being adopted in our developed UQ model. Also, authors developed surrogate modeling-driven Sobol' indices-based global SA to understand associated sensitivity.

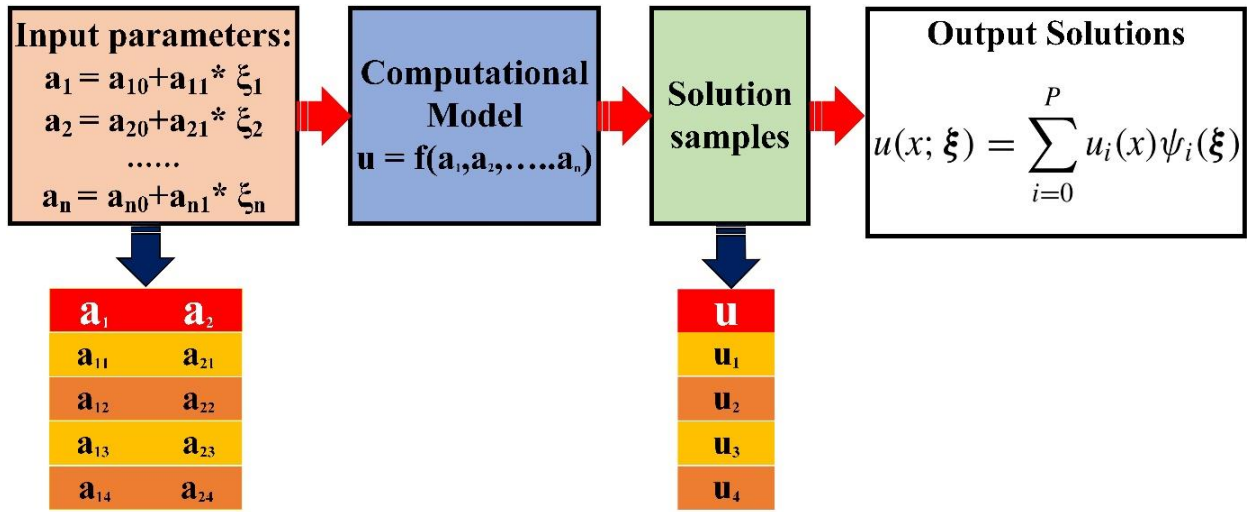

Fig.9. Developed UQ and SA Method (Kumar *et al.*, 2019; Kumar, Alam, Sjöstrand, *et al.*, 2020)

Overall, machine learning and artificial intelligence-driven multi-scale modeling framework for high burnup accident-tolerant fuels for light water-based SMR applications is shown in Fig.10.

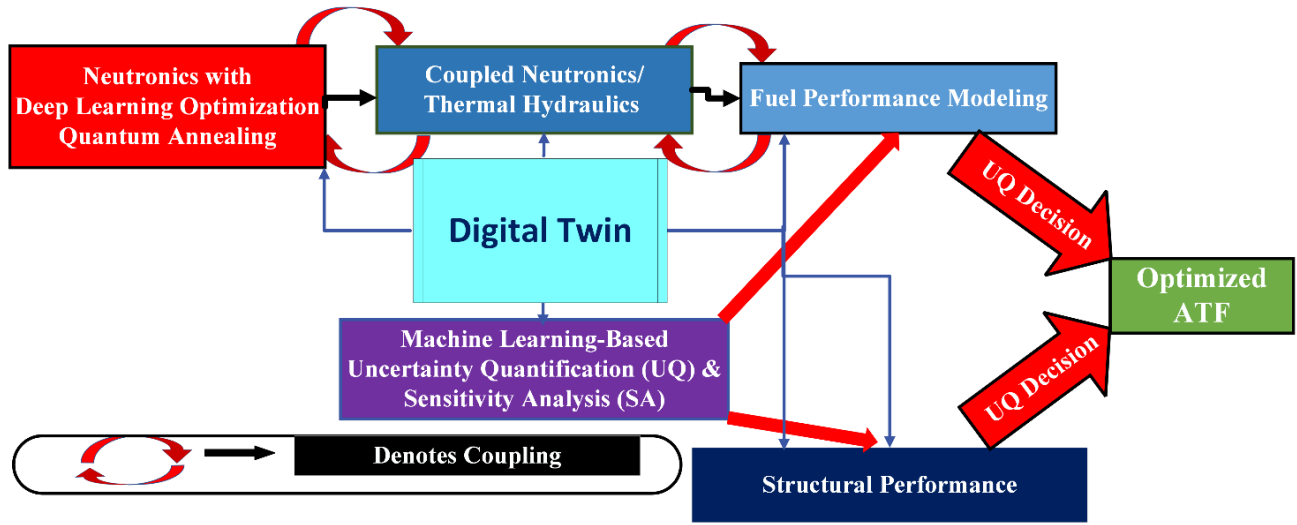

Fig.10. Machine Learning and Artificial Intelligence-Driven Multi-Scale Modeling Framework for High Burnup Accident-Tolerant Fuels for Light Water-Based SMR Applications

## 8. Discussions

From the preceding sections, the first thing that can be identified is the lack of sufficient data on the neutronic and thermohydraulic behavior of ATFs when introduced in LW-SMRs. While most of the studies on ATFs were conducted on the potential use of these fuels in conventional LWRs, very few of the studies considered introducing ATFs in LW-SMRs. Most of the proposed LW-SMR designs propose conventional $UO_2$-Zr fuel system (Aydogan, 2016). And experimental data for SMRs with ATF core is even rarer. Most of the research on the introduction of ATF in a LW-SMR are based on computational tools. This was somewhat expected since SMRs are still on the developmental stage and these reactors are yet to be installed to gather sufficient data. Nevertheless, adequate test facilities for LW-SMRs should be able overcome this data gap.

Another important observation from the literature survey is that AI is yet to be employed for developing high burnup ATFs. The recent studies employed AI in design optimization of reactor core, reactor control and monitoring and failure study, but the use of AI in developing new ATFs for next generation reactors is not observed in the available literature. This may or may not be due to the lack of available data to train AI for accurate predictions since the concept of ATF has come in the limelight just a few decades ago. Still, if AI is utilized to its full potential, it may contribute to the drastic advancement of nuclear fuel technology; like what it has done for other branches of material science such as degradation analysis, nanomaterial analysis, new material discovery, etc. (Wei *et al.*, 2019).

Finally, the literature review indicated that the application of ML and AI in nuclear industry is quite versatile. However, there are certain research domains where AI is yet to be introduced. One such domain is the prediction of thermohydraulic and neutronic safety parameters of a LW-SMR fueled with accident tolerant materials. Lack of research and operational data for high-burnup composite accident-tolerant fuels for LW-based SMR application is a major reason for the absence of the AI-based research in the available literature. If sufficient data may be generated for training, AI can be a strong but reliable computational tool to avoid costly experiments with real-life reactors.

**Conclusion**

Safety concern is the main obstacle to the implementation of nuclear power in many countries of the world. Although the possibility of major accident and radioactive contamination of the surrounding area is very low in the Gen III+ reactors, the fresh wound caused by Fukushima disaster on the minds of millions of people across the globe is a matter of concern to the nuclear power industry. This concern is the driving force for the manufacturers of the NPPs to develop safer reactors for future. A step towards safer Gen IV reactors is the development of accident-tolerant fuels. Because of the superior resistance to degradation of fission product release, these fuels are expected to replace conventional fuels soon. And since the small modular reactors are believed to be the safer and more economic substitutes of their larger versions, the implementation of ATFs in the SMRs. This literature review attempts to present the picture of current trend in the research on ATFs and LW-SMRs. The work also discusses the use of machine learning and artificial intelligence in the field of nuclear engineering and explores the prospect of exploiting this powerful computational tool to accelerate the development of high burnup ATFs. From the literature survey, it is identified that the research focused on the development of ATFs for LW-SMRs are far from realizing the benefits AI because of the lack of sufficient data. This data inadequacy can only be overcome through experiments in test SMR facilities. Further research on high burnup ATFs is required to brough about revolution in the fuel technology of LW-SMRs.